\documentclass[onecolumn,amsmath,amssymb,prd,reprint,longbibliography,floatfix,8pt]{revtex4-1}

\usepackage{natbib}
\usepackage{array}

\usepackage{tabularx} 

\usepackage{mathtools}
\usepackage{cancel}
\usepackage{enumitem}
\usepackage[utf8x]{inputenc}
\usepackage{graphicx}
\usepackage{dcolumn}
\usepackage{bm}
\usepackage[switch]{lineno}
\usepackage{tikz}
\usepackage{multirow}
\usetikzlibrary{arrows.meta}
\usetikzlibrary{arrows,decorations.pathmorphing,backgrounds,positioning,fit,petri}

\begin{document}

\title{Efficient training of generative models from multireference simulations and its application to the design of Dy complexes with large magnetic anisotropy}

\author{Zahra Khatibi}
\author{Lorenzo A. Mariano}
\author{Lion Frangoulis}
\author{Alessandro Lunghi}
\email{lunghia@tcd.ie}
\affiliation{School of Physics, AMBER and CRANN Institute, Trinity College, Dublin 2, Ireland}

\begin{abstract}
\bf Generative machine learning models can potentially provide direct access to novel and relevant portions of the full chemical space, overcoming the cost of systematic sampling. However, the training of these models generally requires a large amount of data, often precluding the use of expensive high-level ab initio simulations for this task. The generation of coordination compounds of Dy with large magnetic anisotropy represents a topical example, where multireference simulations of large molecules are necessary to perform reliable predictions. Here, we show that a semi-supervised chemically-inspired training-by-proxy of generative variational autoencoders can reduce the cost associated with building a training set from multireference simulations by two orders of magnitude. We illustrate the power of this approach by generating 100s of new organic ligands for Dy(III) pentagonal bipyramidal complexes exhibiting record values of magnetic anisotropy, while starting from datasets as small as 1k multireference calculations. This work thus paves the way to the computational generation of molecules as complex coordination compounds with target electronic and magnetic properties.
\end{abstract}

\maketitle

\twocolumngrid

\section{Introduction}
\label{sec:introduction}

Artificial Intelligence (AI), and generative AI in particular, is reshaping the paradigm of materials science by enabling the discovery of novel crystal \cite{ren2022invertible,park2025guiding,duan2025rise, zeni2025generative,de2026generative} and molecular structures \cite{kusner2017grammar,du2024machine,toney2025exploring}. Generative models are highly versatile, and a wide range of architectures has been explored with applications extending well beyond, but not limited to, drug discovery \cite{tong2021generative, qi2024predicting}, catalysis \cite{moran2025ai,balcells2025co,mok2024generative}, sensing \cite{sharma2025fusion,tan2023novo}, and transition-metal complex design \cite{lee2023joint, jin2024partial,strandgaard2025deep}. These include diffusion models \cite{luu2023generative, jin2024partial}, generative adversarial networks \cite{hisama2024theoretical}, reinforcement learning \cite{olivecrona2017molecular,popova2018deep,you2018graph,dodds2024sample,park2025guiding,nguyen2025inverse}, LLM-based architectures \cite{ishida2025large,Lu2025generative}, transformers \cite{chen2023molecular,mok2024generative}, variational autoencoders (VAEs) \cite{gomez2018automatic,jin2018junction,simonovsky2018graphvae,fallani2024inverse,strandgaard2025deep,abeer2026enhancing,dollar2021attention,inukai2025leveraging}.

Despite the increasing number of promising proof-of-concept studies that have recently appeared, generative models have so far been largely applied to chemical problems where either large databases of properties already exist \cite{gomez2018automatic} or where the generation of new molecules can be guided by chemical or physical properties that are relatively inexpensive to compute \cite{abeer2026enhancing}. There are, however, plenty of relevant chemical systems and related properties whose complexity makes it impossible to generate enough data to train generative models, in some cases reaching millions of samples \cite{inukai2025leveraging}. Topical examples are correlated materials, coordination compounds, electronic excited states, and magnetic properties, where expensive multireference simulations are very often needed to get even qualitative predictions. Extending the utility of generative models to such complex scenarios remains a significant challenge and an urgent one to be addressed for the field of machine learning to become a useful and established tool in computational chemistry. \\

In this work, we take on this challenge and demonstrate that generative models can be efficiently trained over expensive multireference properties to conditionally generate complex molecular structures with tailored magnetic properties and electronic excitations. To achieve this goal, we select Dy(III) single-molecule magnets (SMMs) as our test bed. SMMs are often composed of a central ion, either a transition metal or a lanthanide, surrounded by coordinating organic ligands, and exhibit magnetic bistability \cite{zabala2021single}. If properly controlled, such behavior might find applications in spintronics and high-density memory storage \cite{moreno2021measuring}. While many factors ultimately control the magnetization relaxation time underpinning the bistability of SMMs \cite{lunghi2022toward}, some very specific electronic features are at the heart of their magnetic behaviour \cite{rinehart2011exploiting}. When the organic ligands surrounding Dy impart a strong, axial crystal field on the ion's f electrons, then the first eight electronic energy levels of the Dy coordination compound roughly coincide with the possible orientations of a magnetic moment $J=15/2$. More specifically, each energy level is doubly degenerate and goes from a ground state $M_J=\pm 15/2$ up to $M_J=\pm 1/2$. The closer these Kramers Doublets (KDs) are to pure $M_J$ values and the higher their energy gap from the ground state ($\Delta E_{0i}$, $i=1-7$), the larger the magnetic anisotropy and the more robust magnetic bistability is against temperature.

Ab initio simulations based on Complete Active Space Self-Consistent Field (CASSCF) have played a key role in improving the performance of SMMs \cite{lunghi2022computational}, but due to their expensive and delicate nature, most studies target a few molecules at a time, often in response to experiments. Recent approaches aimed at accelerating the screening of SMMs has focused on high-throughput\cite{mariano2024charting, HTpaper} or evolutionary exploration of different ligand environments \cite{frangoulis2025generating}, in search of large axial crystal fields. While promising, these methods might remain constrained by the size and diversity of the available ligand datasets or risk remaining confined to local regions of the chemical space. 

The use of generative models would naturally expand the scope of these recent efforts and possibly provide an unprecedented boost to the discovery of next-generation SMMs. Here, we explore VAE for this task \cite{kingma2013auto,rezende2014stochastic}. VAEs are among the most widely adopted generative models in data science, with successful applications in image, text, and music generation, as well as data augmentation. VAEs compress high-dimensional data into a lower-dimensional continuous latent space described by a probabilistic distribution and reconstruct the input while retaining its essential features. In a chemical context, VAEs enable the generation of novel compounds by introducing controlled stochasticity in the latent space, producing molecules that remain chemically consistent with known structures \cite{gomez2018automatic}. When coupled to predictive models of molecular properties, they allow targeted exploration of chemical space, guiding the design of molecules with desired functionalities. Existing implementations typically rely on graph- or string-based molecular representations, such as SMILES \cite{WeiningerSMILES1980}, which encode composition and connectivity but neglect conformational information. In addition to the large dataset size generally needed to train VAEs, this aspect is a critical limitation for SMMs, as their magnetic properties are incredibly sensitive to even subtle conformational distortions.

\begin{figure}[t!]
    \centering
    \includegraphics[width=\linewidth]{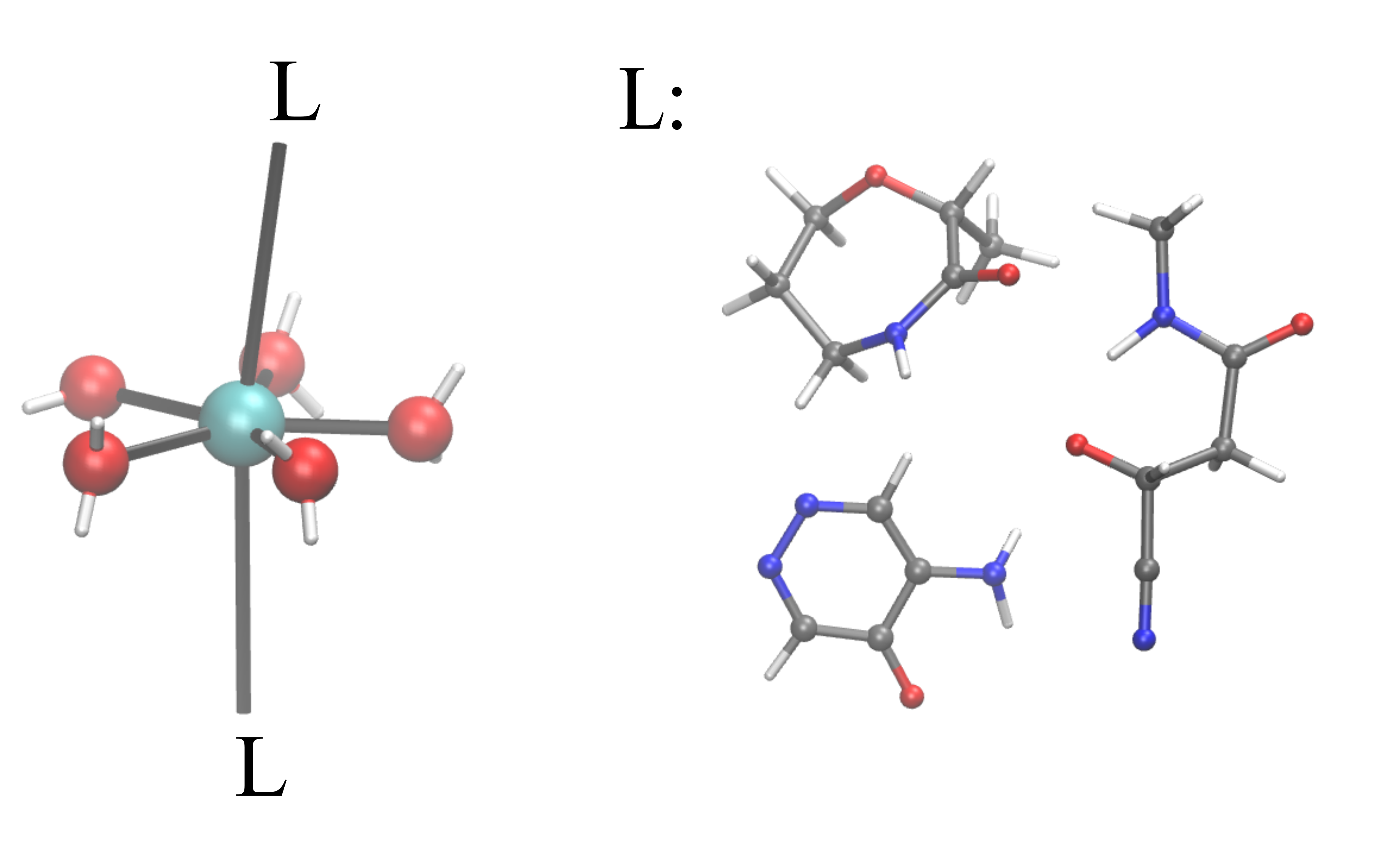}
    \caption{\textbf{Schematic representation of a pentagonal bipyramidal Dy(III) SMM used in this work.} \textbf{Left} The Dy ion (light blue) is coordinated by five water molecules (red= oxygen, white= hydrogen),  and two axial ligands L. \textbf{Right} Selected examples for L (grey=carbon, blue=nitrogen).}
    \label{fig:mol_scheme}
\end{figure}

We solve these challenges with a two-pronged approach: i) we use semi-supervised learning \cite{kingma2014semi,lee2023joint} to train the VAE on a large number of readily available SMILES encoding organic ligands while not requiring all of them to be labelled with molecular properties, and ii) we label SMILES with a chemical property of the sole organic ligands that serve as a proxy for magnetic anisotropy and that can be easily computed with single-reference methods like density functional theory (DFT). This training approach is shown to impart a structure to the VAE's latent space that preserves the locality of molecular structure-property relations going from the trained proxy quantity to magnetic anisotropy without requiring any retraining or fine-tuning. This entirely bypasses the need to carry out large-scale CASSCF simulations to generate a training set, and only requires calculations on the full Dy complex structure, instead of the much smaller organic ligands, to identify the promising areas of the latent space to be sampled. Overall, this training strategy brings down the cost of training a VAE by two orders of magnitude, making it possible to explore demanding chemical problems. Our ultimate model, GAUSS-II (Generative AUtoencoders for State-of-the-art Single-molecule magnets), indeed shows the conditional generation of novel pentagonal bipyramidal Dy(III) complexes (e.g. Fig. \ref{fig:mol_scheme}), one of the most promising classes of SMMs \cite{chen2016symmetry,ding2016approaching,duan2022data}, with record properties starting from just 1k CASSCF calculations. The proposed approach is expected to be generally applicable to coordination compounds and other challenging problems, provided that a suitable proxy property is identified, thus paving the way to the mainstream application of generative models to complex systems and their properties.

\section{Results}
\label{sec:results}

\subsection{The VAE architecture}
\label{sec:just_VAE}

\begin{figure*}[ht]
    \centering
    \includegraphics[width=\linewidth]{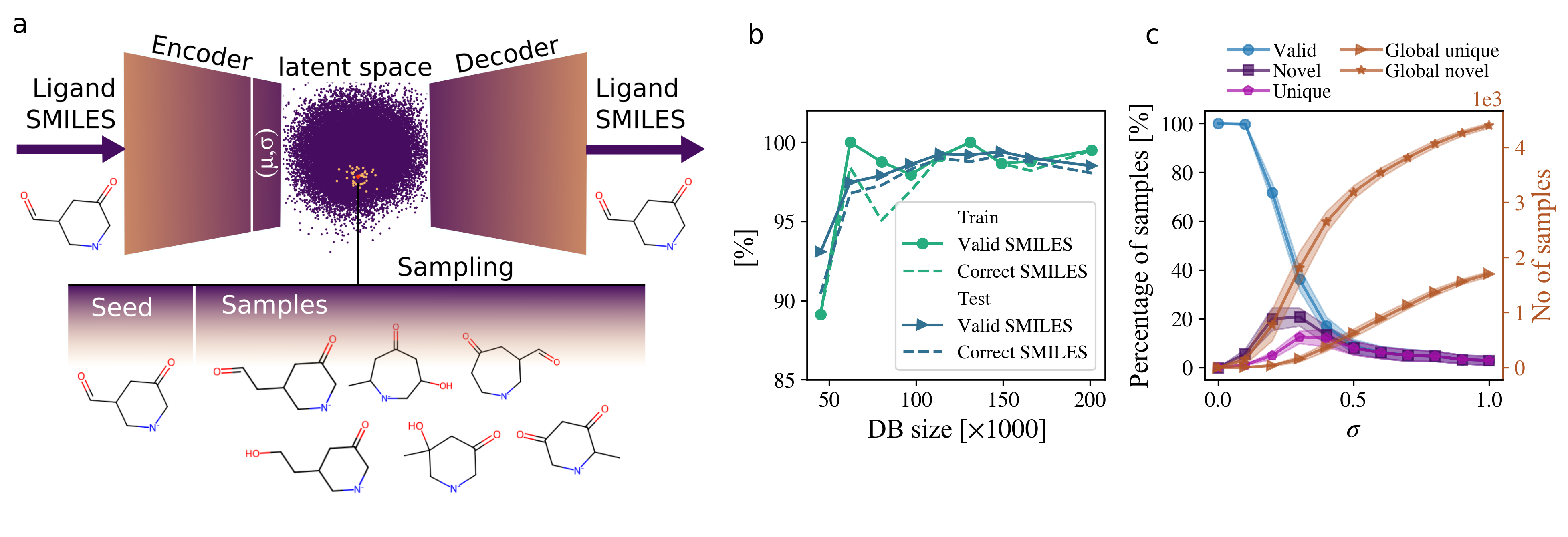}
    \caption{{\bf VAE trained on ligands and its sampling performance.} {\bf a} Schematic representation of a VAE architecture. Input SMILES strings are converted into one-hot encodings and fed into an encoder that maps the training data into a continuous latent distribution. A reparameterization method \cite{kingma2013auto} is then employed to sample the latent space, which is passed to the decoder to reconstruct the input data. After training, a random seed from the latent space can be used with LP sampling to generate multiple samples, many of which correspond to novel molecules. {\bf b} Learning curve of the VAE model, showing reconstruction accuracy as a function of training set size. {\bf c} Sampling rates generated using a single random seed. For each std value, 50 batches of 100 samples were generated and labeled as novel if they were not present in the training set, and unique if they were not repeated within the generated samples at each step. The accumulated novel and unique samples over time are presented as the global variables. Based on the learning curve shown in b, we selected a training set of 114k samples for the sampling experiments.}
    \label{fig:just_VAE}
\end{figure*}

VAEs are composed of two neural networks, the encoder and the decoder, and learn the patterns of data by reducing their dimensionality. As the encoder processes the input and learns the parameters of a distribution that best represents the underlying data structure, the decoder simultaneously learns how to regenerate the input from this encoded distribution. Through this process, data are mapped into a probabilistic distribution known as the latent space, which captures the essential features of the training set (See Fig. \ref{fig:just_VAE}a). Both the encoder and decoder are deep neural networks, typically sharing a similar architecture but in reversed dimensionality. Depending on the nature of the data, the en(de)coder can consist of dense layers, convolutional layers, or recurrent neural networks. However, the final layer of the encoder is always a dense layer that learns the parameters of the latent distribution, its mean, $\mu$, and standard deviation (std), $\sigma$. The arbitrary sampling of the latent space, through a systematic variation of these two parameters, enables the generation of novel look-alike data \cite{kingma2013auto,rezende2014stochastic}. 

The model is trained by minimizing a loss function that includes the reconstruction term, namely the difference between the input and its reconstruction (Loss$_{\rm recon}$), and a regularization term, the Kullback–Leibler divergence (Loss$_{\rm KL}$), which encourages the learnt distributions, defined by the values of $\mu$ and $\sigma$, to follow a normal distribution. This regularization prevents the model from collapsing into sharp distributions with different means, effectively corresponding to distinct points in the latent space rather than the latter being spanned by smooth and continuous distributions. Without regularization, the latent space would develop dead regions, i.e. regions that the decoder cannot process into something meaningful, thus hindering sampling and the generation of novel data. 

Inspired by the work of G{\'o}mez-Bombarelli et al. \cite{gomez2018automatic}, our model is trained on the SMILES string representations of organic molecules. These organic molecules will later be employed to coordinate the Dy ion and generate novel SMMs. After extracting a dictionary of possible SMILES characters from the training set, we construct two-dimensional one-hot encoded tensors with dimensions corresponding to the string length and the total character count. The string length is defined by the longest SMILES in the dataset.
Because SMILES are sequential data with temporal dependencies, such as matching parentheses, closing rings, or maintaining correct atom order, we employ gated recurrent units (GRUs), which are well-suited for processing sequential information \cite{rumelhart1986learning,elman1990finding,karpathy2015rnn}. The details of the encoder and decoder architecture are discussed in Section \ref{sec:methods}. We experimented with different latent space dimensions and found that a 32-dimensional latent space provides the best balance between reconstruction accuracy and novelty across all VAE flavors considered in this work.

For the training of the model, we collated a novel dataset, parsed and curated from the QM9star database \cite{tang2024qm9star}, containing 208k molecules that can potentially act as monodentate ligands. This includes 72k mono-anions, where the charged atom is either nitrogen or oxygen, and approximately 136k neutral ligands. In both cases, nitrogen or oxygen (charged in the anionic species) serves as the coordinating atom. 
For the VAE to correctly encode the SMILES strings, it is essential that the model can identify the coordinating atom in each ligand. Therefore, we used the atom, bond, and charge information from the QM9star database together with the PySMILES software \cite{pysmiles} to regenerate SMILES strings in which the first character corresponds to the coordinating atom. Further details on data processing and the generation of the input data can be found in Sec. \ref{sec:methods}.
For training, we reserved approximately 7k ligands as the test set and, starting from 45k, generated multiple training set sizes from the remaining data. The model was then trained on each subset individually and evaluated to produce the learning curve shown in Fig. \ref{fig:just_VAE}b.
All dataset sizes exhibit excellent reconstruction accuracy, even for the modest dataset size of 45k. The reconstruction score converges rapidly, reaching a plateau at around a training set size of 62k, with the 114k datasets achieving one of the highest accuracies on the test set.

After the training is complete, a latent space sample (\(z^\prime\)) can be generated using a local perturbation (LP) sampling method: A latent vector $z$ (the seed), a chosen std $\sigma$, and a random noise drawn from a normal distribution $\epsilon$ are combined to produce a sample latent vector according to \(z^\prime=z+\sigma \epsilon\). 
The newly generated latent samples are then passed to the decoder for reconstruction. We next use the RDKit software package to identify valid SMILES strings, i.e. strings that correspond to chemically valid molecules  \cite{rdkit}. From these, we extract the novel molecules. 
\begin{figure*}[t]
    \centering
    \includegraphics[width=\linewidth]{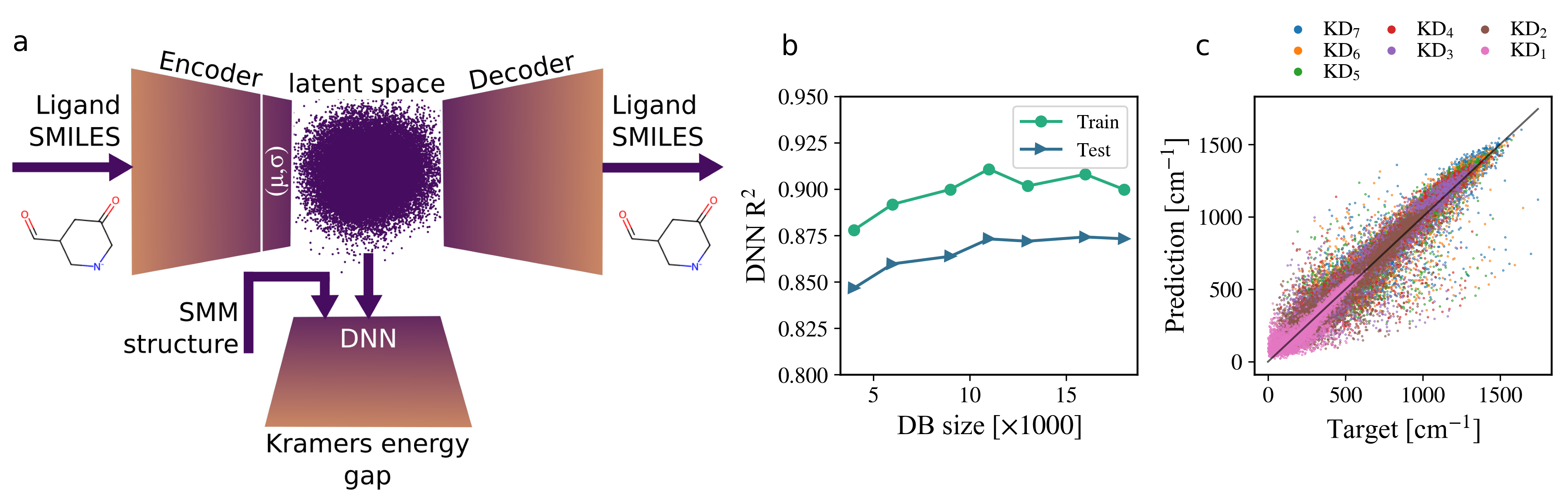}
    \caption{{\bf VAE trained on ligands and their structural features.} {\bf a} Architecture of a VAE model with guided sampling using structural information. Structural features of the associated SMM after DFT optimization, together with the corresponding latent vector, are fed into a DNN to predict the KD energy gaps. {\bf b} Learning curve of the model shown in a, illustrating the convergence of the DNN R$^2$ score as the training set size increases. The R$^2$  score is defined as the ratio between the squared distance of the predictions from the target values of all KD energy gaps and the squared distance of the target values from their mean. This results in an R$^2$ vector of length 7, which is averaged for presentation in b. {\bf c} Parity plot of all KD energy gaps for a model trained on a dataset in which 11k samples are labeled. }
    \label{fig:VAE_xyz}
\end{figure*}
Sampling results for a random seed from the 114k dataset are shown in Fig. \ref{fig:just_VAE}(c). For this analysis, we generated 50 batches of 100 samples for each std value in the range $[0,1]$. The novelty of the generated samples was evaluated by comparing them against the training set. In addition, each unique sample was counted only once in the unique subset, as some samples may be generated repeatedly during the process.
As expected, the novelty of the samples increases as the sampling region moves farther away from the seed. Conversely, the percentage of valid SMILES decreases from 100\% at zero std, where all samples are identical to the seed, and follows a bell-shaped trend, approaching zero near a std of 1. This behavior is consistent with the latent space encoding, where a single point is mapped to a normal distribution. Among the valid strings, a smaller fraction is novel and an even smaller fraction is unique, with both quantities reaching their maximum at a std of approximately 0.3.
On the right axis of the plot, we report the accumulated number of novel and unique samples obtained after approximately 55k sampling attempts from a single random seed. We note that achieving $\sim$2k unique samples takes a few minutes on a standard desktop computer, highlighting the efficiency and generative power of the model.

\subsection{Semi-supervised training and property-guided generation of SMMs}
\label{sec:VAE_xyz}

Now that we have a VAE able to generate novel organic ligands, we want to bias the generation towards ligands that would impart the maximum crystal field splitting to the ground state multiplet of a Dy(III) ion to produce large magnetic anisotropy. As anticipated, here we focus on the promising class of Dy(III) pentagonal bipyramidal SMMs. This class of compounds represents an ideal platform for the design of SMMs, given their promising prospects for experimental realization and technological advancement \cite{Liu_2016, Ding_2016, gupta2016air, Woodruff2013}. In particular, we pick [Dy(H$_2$O)$_5$L$_2$]$^{+}$, with L=($^t$BuPO(NH$^i$Pr)$_2$), as a template for our study \cite{gupta2016air}. In this molecule, Dy(III) is coordinated by five planar equatorial water molecules, as depicted in Fig. \ref{fig:mol_scheme}, and two identical axial ligands. 
This compound is particularly interesting to us, not only because it exhibits a large magnetic anisotropy, but also because it was found to be air-stable, an ideal property for newly designed SMMs to inherit. Moreover, as our target is to generate novel axial ligands, the presence of small water molecules in the equatorial plane decreases possible issues arising from excessive steric hindrance.

To achieve the generation of ligands that maximise the template's magnetic anisotropy, we extend the VAE architecture by integrating an additional deep neural network (DNN) that is trained directly on the latent space features to predict the target property, the KDs energy gaps in our case. An extra mean-squared error term (Loss$_{\rm DNN}$) is added to the total loss function to ensure that the model learns both to reconstruct the data and to predict its associated property. This seemingly trivial extension of the VAE's architecture comes with two significant challenges: i) accurately predicting KDs energy gaps, and ii) generating a large-enough training set.

Starting from the first point, we note that a naive implementation of this architecture to generate SMMs with targeted KD gaps would involve labeling the ligands' SMILES with the KDs energy values for the associated Dy complex. However, this approach does not work in practice and leads to poor results. The model descriptors, i.e. SMILES, contain only minimal information about the ligand, capturing composition and topology but lacking any explicit structural or geometrical detail. This is a particularly striking limitation in the context of magnetic properties as the gaps between KDs are strongly influenced by the molecular conformation, which plays a crucial role in shaping the charge density and crystal field splitting. As a result, SMILES alone are insufficient for the training of the predictive component of the model. See the Supplementary Material for a more detailed discussion. To address this issue, the features input to the DNN are extended with structural information extracted from the DFT geometry optimization of the full Dy complex, as represented in Fig. \ref{fig:VAE_xyz}a. The structural input, a six-dimensional vector, is inspired by our previous work \cite{frangoulis2025generating}, and includes the atomic number of the coordinating atom in the ligand, the mean plane parameter (MPP), defined as the sum of distances of in-plane water molecules from a plane fitted to these molecules, minimizing the deviations, and the spread of distances parameter (SDP), which is the difference between the maximum and minimum distances of the water molecules from the fitted plane \cite{lu2021two}. Finally, the vector also includes the angle formed between the coordinating atoms of the axial ligands and the Dy ion. These key structural features have been identified as the most effective descriptors, exhibiting a strong correlation with KD splittings in local D$_{\mathrm{5h}}$ symmetry systems \cite{frangoulis2025generating}.

Although this choice of structural features is likely going to lead to good predictions of KDs energies \cite{frangoulis2025generating}, this strategy would require DFT optimizations and CASSCF simulations of the KDs energy gaps for the entire dataset, i.e. the full 208k SMILES. Even for the minimum training set size of 114k structures needed to achieve a good reconstruction rate, such a volume of calculations is computationally prohibitive, which leads us to the discussion of the second challenge introduced earlier. 
\begin{figure}[t]
    \centering
    \includegraphics[width=\linewidth]{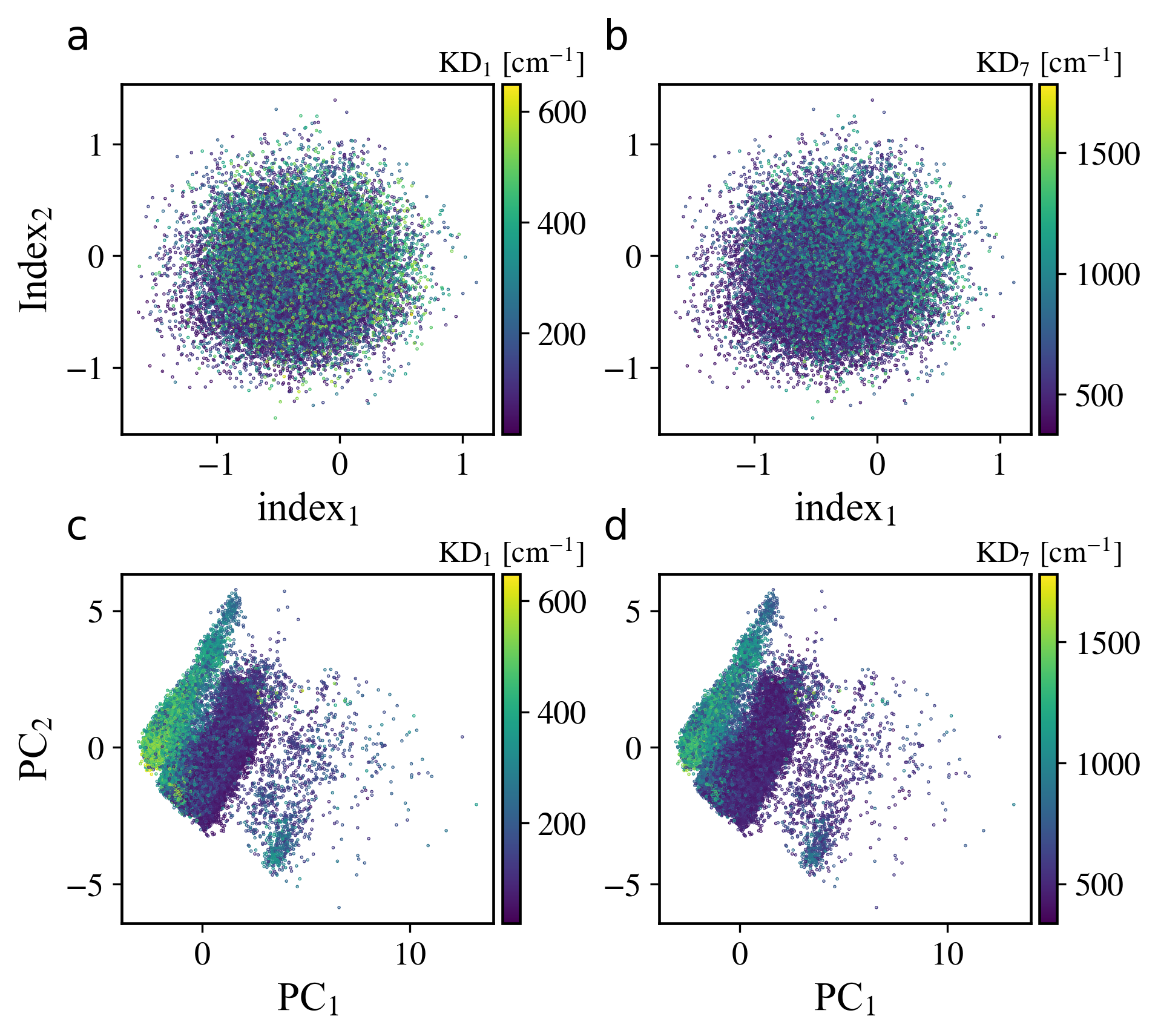}
    \caption{{\bf Latent space ordered by structural features.} {\bf a, b} The span of the latent space along two randomly selected dimensions of the 32-dimensional latent vector. The scatter plots are colored according to the predicted first and last Kramers energy gaps, obtained from a DNN that takes as input the latent vector concatenated with structural features of the fully relaxed SMM, such as the SDP and MPP descriptors. {\bf c, d} show the PCA of the latent space in 2D. The model was trained on a dataset of 129k molecules, of which only 18k ligands were labeled with Kramers energy gaps. The plots shown here correspond to the full set of 23k labeled molecules, including the 18k used during training.}
    \label{fig:VAE_xyz_latent_distr}
\end{figure}
We employ semi-supervised learning \cite{kingma2014semi,lee2023joint} to circumnavigate this bottleneck and only partially label the entire SMILES dataset. The structural information used for training was thus only evaluated for a subset of 23k ligands drawn from the full 208k dataset. This subset was produced in Ref. \cite{HTpaper} via random sampling from a PCA mapping of the ligands’ bispectrum components, to maximize chemical diversity. In the same work, the selected ligands were also assembled into Dy(III) mononuclear complexes using the same template as in Fig. \ref{fig:mol_scheme}, geometrically optimized using DFT, and finally screened with CASSCF to compute the KDs energy gaps.

To evaluate the model performance and predictions accuracy, we reserved a fixed subset of size 4k from the 23k labeled dataset as a test set and constructed multiple training set sizes from the remaining data to generate a learning curve. Since the objective here is to assess the property-prediction capability of the model, we disabled the decoder by removing the reconstruction loss and focused exclusively on minimizing the DNN prediction loss. The results of this analysis are presented in Fig. \ref{fig:VAE_xyz}b.
We measured the model performance by evaluating the DNN R$^2$ score. Note that the targets, and consequently the predictions, are 7-dimensional vectors encompassing all seven Kramers gap energies. Accordingly, the R$^2$  metric is also a 7-dimensional vector, with each component corresponding to the prediction accuracy of one Kramers gap. For clarity and convenience, we report the R$^2$ score averaged over all seven levels. The results indicate excellent predictive performance and strong generalization even for a minimal training set size of 4k, although the R$^2$ score converges at a larger training set size of approximately 11k. 
The parity plot for this training set size, shown in Fig. \ref{fig:VAE_xyz}c, further demonstrates the accuracy of the DNN in predicting the KD energy gaps.

The simultaneous training of the VAE's reconstruction and property predictions forces the model, GAUSS-I, to place molecules with similar properties close to one another in the latent space. The resulting latent space, which follows a normal distribution, is shown in Fig. \ref{fig:VAE_xyz_latent_distr}a,b, where two randomly selected indices of the 32-dimensional latent vector are used to visualize the distribution.
To further appreciate the ordering of the latent space, we performed a PCA of the latent distribution, as implemented in Scikit-learn \cite{scikit-learn}, and projected it onto the two principal components. The latent vectors are colour-coded according to the energy gap between the ground state KD and the first excited KD (Fig. \ref{fig:VAE_xyz_latent_distr}c) and the last KD (Fig. \ref{fig:VAE_xyz_latent_distr}d), respectively. A complete PCA analysis of the latent space across all KDs energy gaps is provided in the Supplementary Material. 
\begin{figure}[t]
    \centering
    \includegraphics[width=\linewidth]{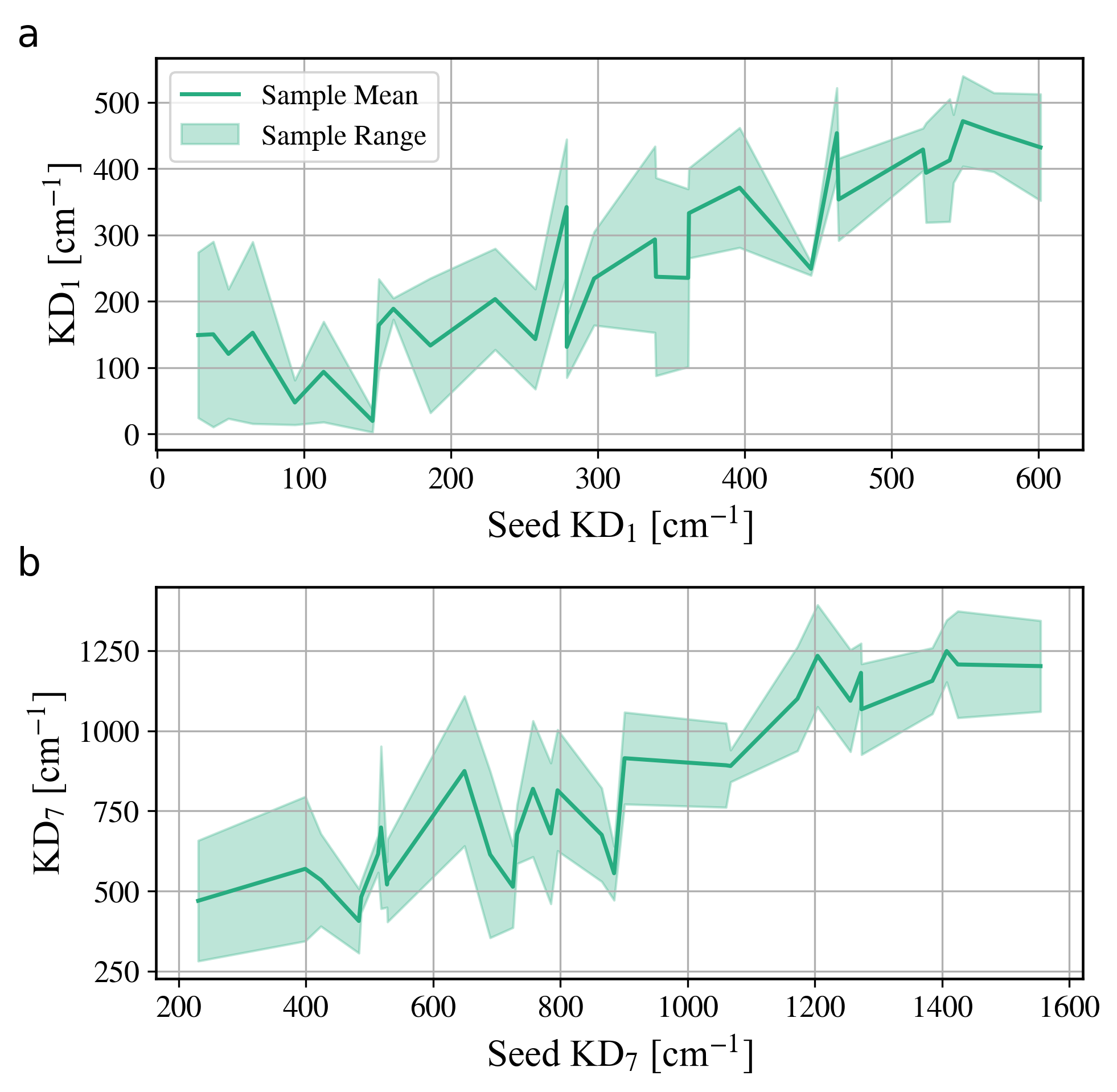}
    \caption{{\bf Kramers energy gaps of random seeds and their samples.} The {\bf a} first and {\bf b} last KD energy gaps of the generated samples versus those of the corresponding seeds. The target KD gaps from the 23k dataset were divided into six bins, from which six batches of random seeds were selected. Each batch contained 10 unique samples that were generated using a scaled std of up to 0.2. The sampling procedure was repeated five times, yielding a total of 300 generated samples. The generated samples were subsequently assembled into Dy(III) SMMs, structurally optimized, and their Kramers gap energies were evaluated using CASSCF calculations.}
    \label{fig:VAE_xyz_sample_KD}
\end{figure}
Having access to latent regions encoded with different KDs energy gaps, we can now select specific ligands from preferred regions, the sampling seeds, to generate new organic ligands, which, according to the model, would lead to SMMs with similar KDs energy gaps. 

To test the model's performance over this conditional generative task, we divided the total span of KDs energy gaps for the training set into several bins and randomly selected multiple seeds from each. These seeds were then used to sample the local latent space through the LP method discussed in Sec. \ref{sec:just_VAE}. 
To generate samples sufficiently close to the seeds, we initially used a small std of 0.1 and performed 4000 LP sampling steps to acquire 10 unique samples. Since the novelty and consequently the uniqueness of the samples is very low at such a small std (see Fig. \ref{fig:just_VAE}c), and 10 unique samples might not be obtained in this initial sampling, we allowed additional sampling steps with an increased std of 0.2 each to reach the target of 10 unique strings.
The successfully generated unique SMILES were then individually assembled as the axial ligands of the Dy template of Fig. \ref{fig:mol_scheme} using the MolSimplify package \cite{molSimplify}. This is then followed by DFT optimization and CASSCF calculations to extract the KDs energy gaps. The results of this analysis are presented in Fig. \ref{fig:VAE_xyz_sample_KD}, where the calculated first and last KDs energy gaps of the generated samples are shown as a function of the corresponding seed values. For both the first and last KD gaps, the generated samples follow the expected trend. Interestingly, the model shows better fidelity in the mid-range of energy values than in the low- and high-energy regimes, possibly hinting at an effect of a reduced number of molecules in the training set for areas at the boundary of the property distribution.

\subsection{Proxy-training and property-guided generation of SMMs}

\begin{figure*}[ht]
    \centering
    \includegraphics[width=\linewidth]{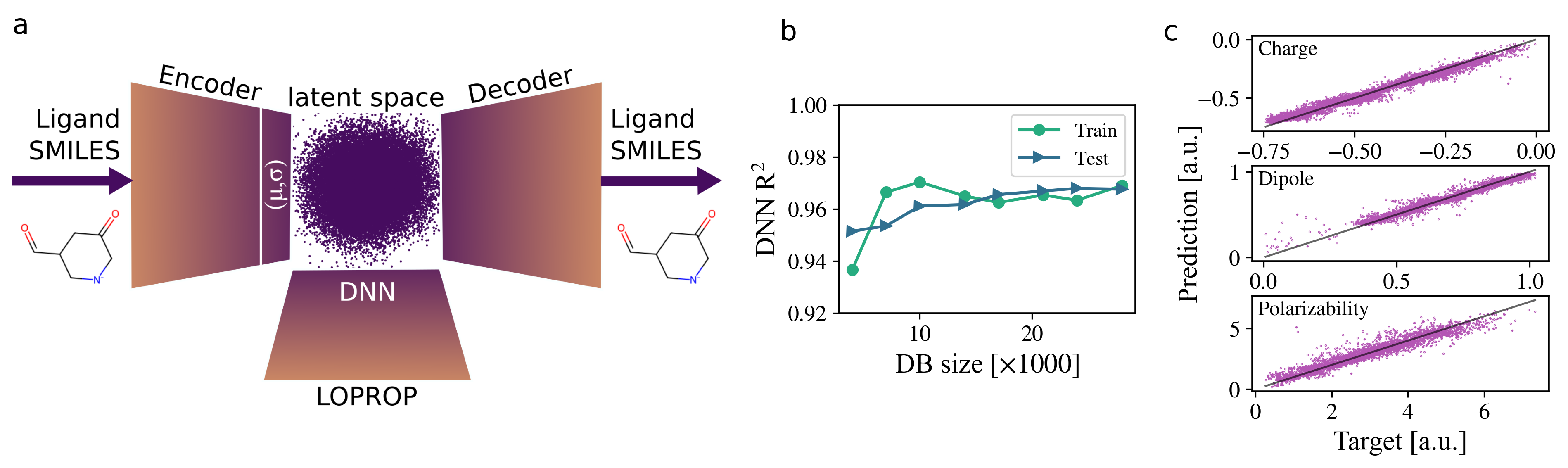}
    \caption{{\bf VAE trained on ligands and their LoProp.} {\bf a} Schematic representation of the VAE model guided by LoProps, which steer the search and sampling within the latent space. A subset of molecules labeled with LoProps is passed to a DNN during training. In addition, a portion of the labeled set contains the first Kramers doublet energy gap which is used for cross-mapping. {\bf b} Learning curve of the model, showing the DNN R$^2$ score as a function of training set size. The training converges rapidly and remains highly efficient even for a modest dataset of 4k samples. {\bf c} Parity plots (DNN predictions versus target values) for a model trained on 17k samples, the smallest dataset size achieving near-converged R$^2$ performance. The connecting atom type is also among the LoProp target labels; however, since it is binary (oxygen or nitrogen), its parity plot is not shown here.}
    \label{fig:VAE_loprop}
\end{figure*}

So far, we have shown the performance of a generative model that can successfully generate novel organic ligands that, once they occupy the axial positions of a Dy(III) pentagonal bipyramidal complex with planar water molecules, lead to the targeted value of magnetic anisotropy. Despite succeeding at this generative task, the training of such a model requires in the order of $\sim$10k labeled samples. This is an important reduction compared to a fully labeled dataset, but still requires a number of DFT geometry optimizations and CASSCF calculations which is far from routine and entails substantial computational cost. Here, we explore a solution to this challenge by avoiding training the model directly on magnetic properties and instead relying on proxy properties that can be computed at significantly lower costs. The problem then reduces to identifying suitable proxy descriptors. The quantum theory of coordination bonds and crystal field in f elements is now very well understood. While both electrostatic and covalent effects are operative \cite{briganti2019covalency}, the crystal field splitting is inherently local in nature and will largely be determined by the chemical nature of the binding atoms and their position \cite{lunghi2017intra}. Here, we select a simple recipe to represent the chemical nature of a ligand's binding atom: the atomic number, the atomic partial charge, the atomic dipole moment, and the atomic polarizability. All these quantities can be extracted from DFT calculations with the LoProp scheme \cite{gagliardi2004local}. This new set of local properties is used for the training of the DNN within the VAE, as done previously for KDs energy gaps, while all subsequent steps are identical to those of the previously discussed model in Sec. \ref{sec:VAE_xyz}, including LP sampling and validation of the generated samples through CASSCF calculations. We then generated and curated a dataset of approximately 35k ligands with their LoProp properties evaluated using MOLCAS \cite{aquilante2020modern}. The molecules included in this dataset were once again selected through a random sampling of a PCA projection of the bispectrum components of the original 208k-ligand set. This dataset fully includes the 23k molecules used in the previous section for which KDs gaps are available.

\begin{figure*}[t]
    \centering
    \includegraphics[width=\linewidth]{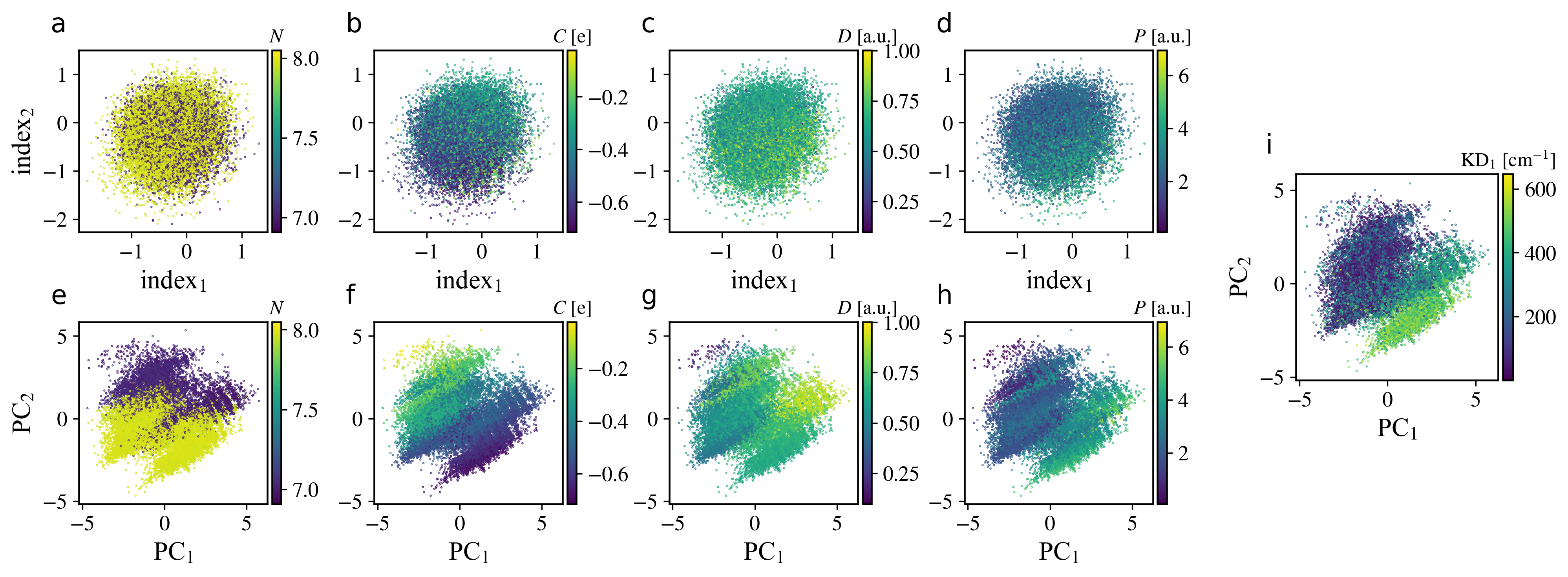}
    \caption{{\bf Latent space ordered by LoProp.} {\bf a-d} The latent space ordered according to LoProp values learned through a DNN that receives the latent vector as input. The axes correspond to two randomly selected dimensions of the 32-dimensional latent space. {\bf e-h} show the 2D PCA projection of the latent space. The coloring in all subplots a-h is based on the predicted LoProp values. The model was trained on a dataset of 114k ligands, of which approximately 29k have associated LoProp values. The data presented here comprise 23k ligands, all of which have LoProp values, with a subset included in the training set.
    {\bf i} PCA of the latent space colored according to the first Kramers gap energy for each latent vector. The cross-comparison of the latent mappings in a–h and i enables sampling from different regions of the latent space for targeted repopulation of the latent and chemical space.}
    \label{fig:VAE_LOPROP_latent_distr}
\end{figure*}

To examine the model performance, we reserved a fixed subset of the data, 7k, as a test set and generated multiple training set sizes from the remaining ligands to construct a learning curve. The results are shown in Fig. \ref{fig:VAE_loprop}b, where we observe fast convergence and an excellent R$^2$ score for the DNN, even for a training set as small as 4k ligands. The high R$^2$ score is further reflected in the parity plots comparing the predicted and target LoProp values, shown in Fig. \ref{fig:VAE_loprop}c. Similarly to what was observed in the previous section, the training of the VAE and subsequent PCA enables the visualisation of the ordered structure of the latent space. This behavior is clearly illustrated in Fig. \ref{fig:VAE_LOPROP_latent_distr}e-h, where we present the PCA projections of the latent space learned by the VAE and colored by the different LoProp values. 

We are now left to map these proxy properties to our target KDs energy gaps. This is, in principle, far from trivial. As we observed in the previous section, attempts to train a regressor that maps SMILES to our target quantities did not lead to satisfactory results, and the introduction of LoProp properties alone only marginally improves the training outcome. However, we here show that this further step is, in fact, not needed. In Fig. \ref{fig:VAE_LOPROP_latent_distr}i, we report the latent space PCA color-coded by the KDs energy gaps and demonstrate that the order imparted to the latent space by the training over LoProp properties naturally extends to KDs energy gaps, making it ready to be sampled for the generative task without requiring any further training or fine-tuning. The cross-comparison of the different latent-space encodings reveals clear structure–property relationships, supporting the fact that the mapping is physically and chemically justified and not a coincidence. For instance, we observe that complexes coordinated by negatively charged oxygen ligands with the largest local charge account for the majority of molecules with large KDs energy gaps, an observation that is fully consistent with chemical intuition and trends previously identified through direct dataset analysis in Ref. \cite{HTpaper}. 

\begin{figure}[t]
    \centering
    \includegraphics[width=\linewidth]{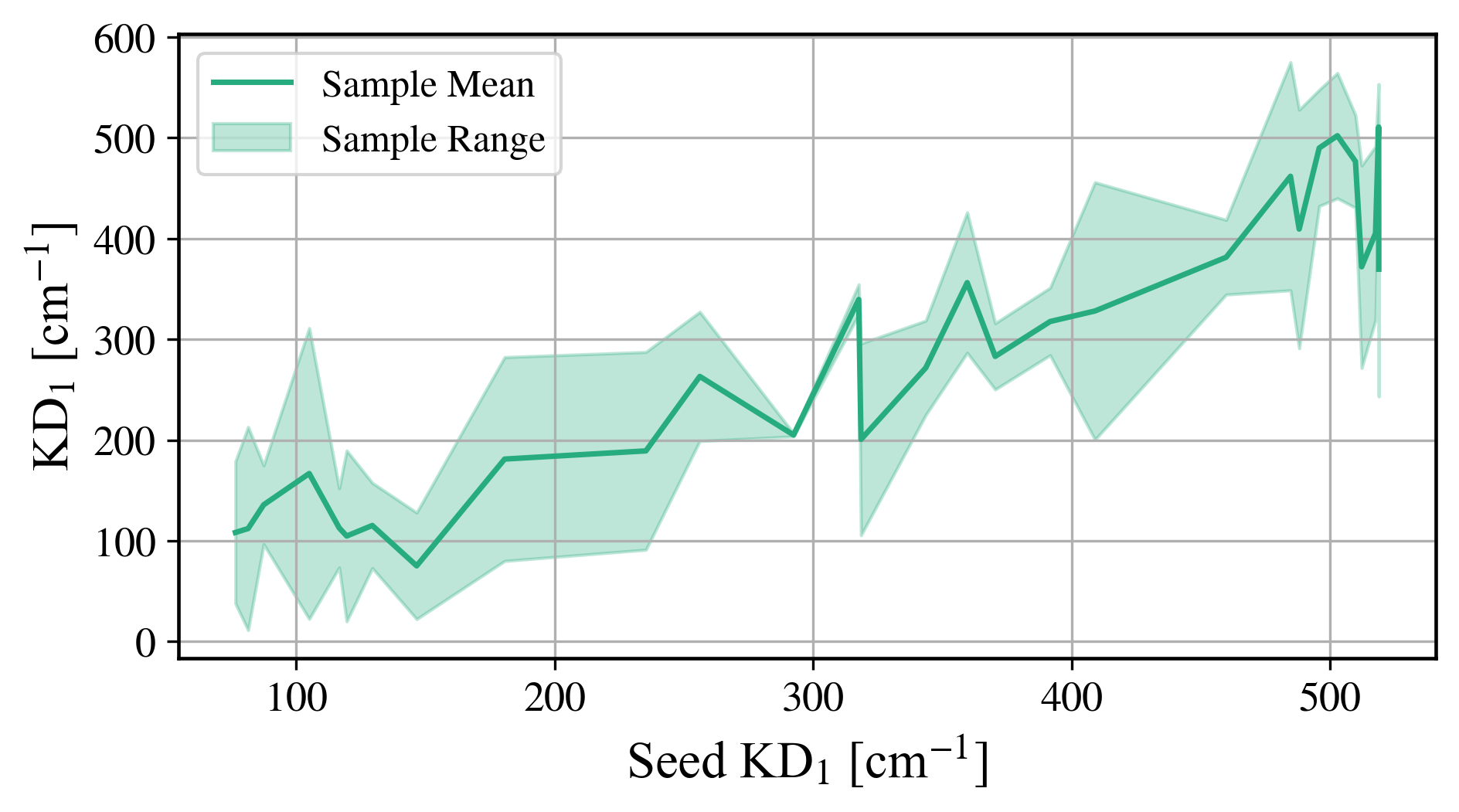}
    \caption{{\bf The first KD energy gap of random seeds and their corresponding samples.} Using a strategy similar to that shown in Fig. \ref{fig:VAE_xyz_sample_KD}, we divided the target KD values into six bins and selected five seeds from each bin. Using the LoProp values of the seeds, the corresponding latent vectors were identified and used for sampling. For each seed, 10 samples were generated within the std of 0.2. All generated samples were then assembled into Dy(III) SMMs, geometrically optimized, and their Kramers gap energies were evaluated.}
    \label{fig:VAE_loprop_sample_KD}
\end{figure}

By leveraging this two-stage mapping of the latent space, we can identify, sample, and repopulate regions of chemical space corresponding to molecules with specific target KDs energy gaps. For this purpose, we adopted a strategy similar to that described in Sec. \ref{sec:VAE_xyz}, i.e. the KD energy range was divided into six bins, from which multiple seeds were randomly drawn in batches. For each seed, ten samples were generated using the LP method with a std of at least 0.1. Finally, the generated SMILES were used to assemble the corresponding Dy complexes using the MolSimplify package. These structures were then subjected to DFT optimization and CASSCF calculations to validate the results, i.e. to assess whether the properties of the generated samples followed those of their respective seeds. The results of this analysis are shown in Fig. \ref{fig:VAE_loprop_sample_KD}. 
In comparison to the previous VAE model trained directly on the KDs energy gaps, this new model exhibits a noticeably improved performance (See Fig. \ref{fig:VAE_xyz}). With the exception of a few seeds in the high–energy regime, the generated samples closely follow the trends of their corresponding seeds across all energy domains.
To compare the sampling performance of the two proposed models, we define a metric similar to the R$^2$ score, in which the KD splittings of the generated samples are compared against those of their corresponding seed molecules. Using this metric, we find that the present variant scoring R$^2$ of 0.82 outperforms the former by 17\%, even though the optimization is guided by proxy descriptors rather than the direct KDs energy gaps.

\begin{figure}[t]
    \centering    \includegraphics[width=\linewidth]{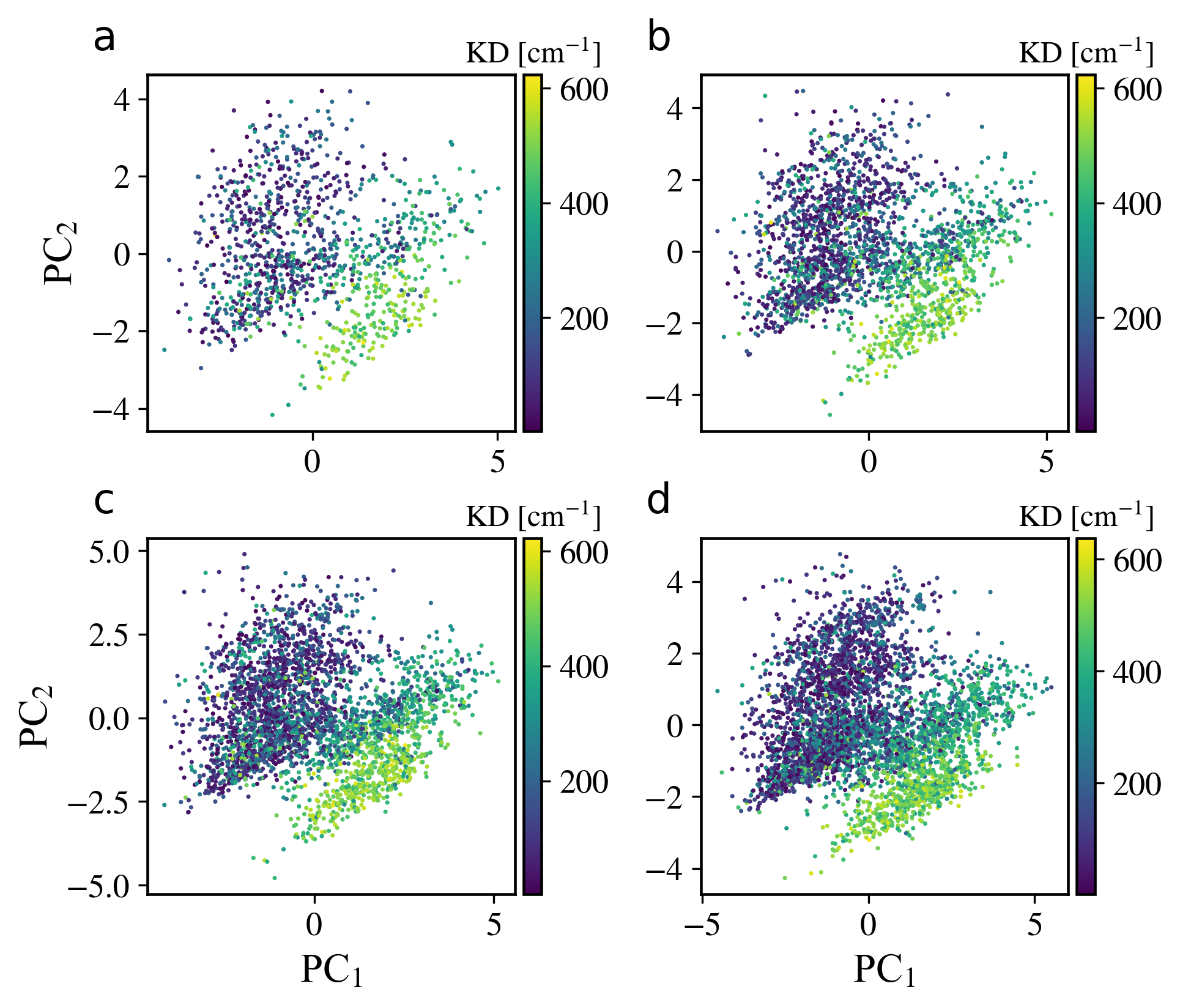}
    \caption{{\bf Latent space mapping with least DB size.} PCA projections of the latent space obtained using {\bf a} 1k, {\bf b} 2k, {\bf c} 3k, and {\bf d} 4k data points, color-coded by the first Kramers splitting energy. Distinct energy regimes begin to emerge already with as few as 2k data points, highlighting the efficiency of the latent representation learned by the model. }
    \label{fig:KD_distr_progress}
\end{figure}

\begin{figure}[t]
    \centering    
    \includegraphics[width=\linewidth]{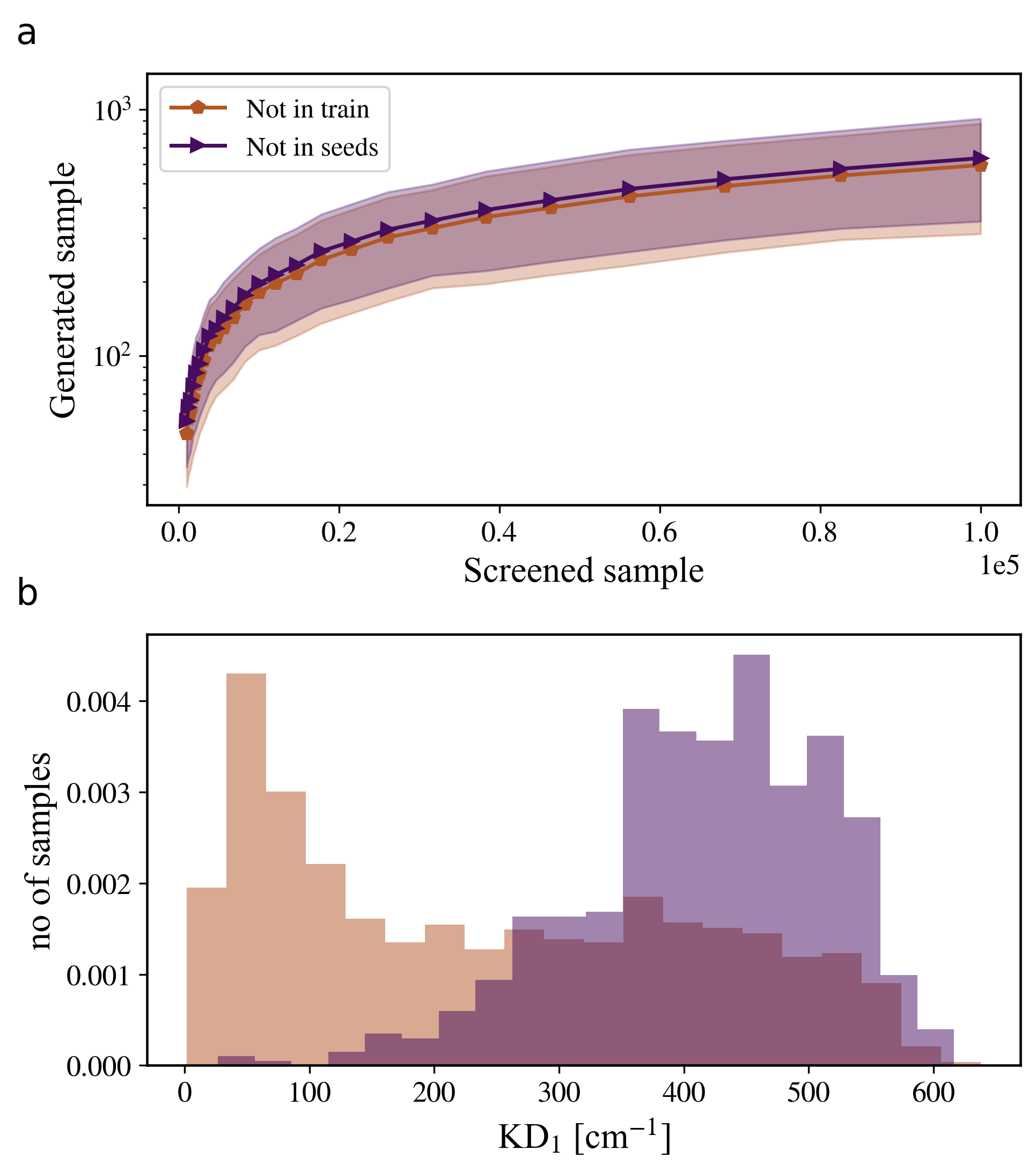}
    \caption{{\bf Targeted repopulation of chemical space.} \textbf{a} Distribution of unique samples versus the total number of screened samples. The solid lines represent the mean number of samples across the 16 seeds, obtained by generating samples independently for each seed and averaging the results. The shaded regions indicate the corresponding range of sample counts. Starting from the top 16 performing seeds selected from the 1k labeled dataset presented in Fig. \ref{fig:KD_distr_progress}, at each step the screened samples were generated using the LP method with \(\sigma=0.3\). The generated samples were then validated using RDKit. The valid strings were compared to the training set and labeled as unique if they did not belong to it. They were further compared to the 1k seeds to evaluate their uniqueness relative to the initial set. \textbf{b} From the generated samples that did not belong to either the training set or the 1k seeds, 800 were selected and validated using CASSCF calculations. A total of 684 converged successfully, and their first \(K_D\) values are plotted against those of the 1k seeds.}
    \label{fig:stats_std}
\end{figure}

Finally, we want to highlight the effectiveness of this ultimate model, GAUSS-II, and the remarkable efficiency of this framework. We randomly selected between 1k and 4k molecules from the training set labeled with KDs energies and examined the distribution of the first KD energy gap (See Fig.\ref{fig:KD_distr_progress}). Strikingly, clear clustering and the emergence of distinct energy regimes are already evident with as few as 1k labeled samples. By 4k, the latent-space organization closely mirrors that obtained using the full set of 23k labeled molecules. This result demonstrates that a minimally labeled dataset is sufficient to individuate promising areas of the latent space and enable the targeted generation of ligands leading to any desired value of KDs energy gaps.  
To further support this point, we identified the top 16 performing seeds from the restricted 1k labeled dataset, which was randomly drawn from the larger 23k labeled dataset and is presented in Fig. \ref{fig:KD_distr_progress}a. Using the LP sampling method with a std of 0.3 we set out to explore i) how many unique novel molecules the model can generate, ii) their KDs energy gaps distribution. Fig. \ref{fig:stats_std}a reports the number of novel and unique molecules generated, both with respect to the total SMILES training set and the 1k molecules used to map the latent space, as a function of the number of samples around a seed. Despite generating a new molecule through the LP sampling is a statistically unlikely event, sampling itself is not a computationally demanding task and hundreds of novel and unique molecules can readily be obtained from each seed. From this unique set of generated molecules, we selected 800 ligands, assembled the corresponding Dy complexes, and evaluated their KD energy gaps using CASSCF calculations. These results are compared to those of the initial 1k labeled set in Fig. \ref{fig:stats_std}b. We observe a clear positive energy shift: aside from a few outliers, the first KD energy gap distribution of the generated set closely follows that of the top 16 seeds. This further confirms the remarkable effectiveness of the model in achieving targeted repopulation of chemical space using a minimally sized labeled dataset of only 1k samples, substantially improving upon conventional high-throughput molecular design strategies for SMMs.

\section{Discussion and Conclusion}
\label{sec:discussion}
Generative AI is posited to change the way we approach the design of new materials. These approaches hold the promise for an accelerated discovery of new materials and molecules by enabling the exploration of a virtually infinite chemical space by learning its underlying features and linking them to molecular properties. However, despite the promise, the success of these models often relies on the availability of large training datasets. This is true for many architectures, particularly more complex NLP-inspired models, which may require millions of data points \cite{inukai2025leveraging}. Such data demands represent a significant limitation for the design of materials with even moderately computationally expensive target properties. 

Our VAE-based generative framework demonstrates exceptional capability in addressing this challenge for a prototypically complex class of molecules such as Dy SMMs. Operating in a semi-supervised regime, the model is trained on a large corpus of unlabeled ligand strings to learn the underlying chemical feature space, while only a subset of the training data is labeled. Moreover, rather than training on full SMM structures, the model operates at the ligand level, learning from compact string-based descriptors that can be readily extracted from online chemical databases. This dramatically lowers the structural and computational complexity of the problem. Furthermore, instead of relying on expensive KDs energies for supervision, the model is trained on significantly less costly proxy properties of the \textit{ligand}, reducing the dependence on high-level ab initio multireference calculations. Ultimately, a very small fraction of samples, approximately 1k, needs to undergo high-level magnetic property evaluation to establish the cross-mapping between the learned latent representations and the final magnetic properties. This data-efficient design enables the targeted generation of chemically realistic, property-oriented SMM candidates while drastically reducing the computational overhead typically associated with magnetic materials discovery.

Besides the specific importance of our results for the field of molecular magnetism, the underlying philosophy of our approach, i.e. semi-supervised learning plus training-by-proxy, can be readily exported to several other classes of molecules and properties. First and foremost, our method directly tackles the vivid challenge of adapting generative models, and machine learning models in general, to coordination compounds \cite{lee2023joint,jin2024partial,briganti2025machine,strandgaard2025deep,frangoulis2025generating}. Despite their high relevance for many fields, this class of molecules is often neglected in mainstream methodological efforts in favour of organic molecules. Our contribution to this emerging field provides strong evidence that the conditional generation of novel coordination compounds is possible. In addition to this, the present results provide a practical guide to reduce the computational cost associated with the generation of training sets labeled with the results of multirefence simulations or other high-end electronic structure methods. Moreover, to the best of our knowledge, the application of our method to SMMs effectively marks the first application of generative models to both molecular magnetic properties and molecular excited states, paving the way to the explanation of electronic properties and excitations in other classes of molecules.

Although the presented model performs extremely well in generating novel SMMs, it can still benefit from further improvement. The current implementation operates on SMILES representations, which are known to have intrinsic limitations, as they are highly sensitive to the exact ordering of characters in the string. This becomes particularly critical in generative models such as VAEs, where molecules are produced character by character and many intermediate samples may be invalid, as illustrated in Fig. \ref{fig:just_VAE}c. One promising strategy to address this limitation is to adopt a graph-based molecular representation. Instead of treating individual atoms as nodes, higher-level chemically valid substructures, such as aromatic rings or functional branches, can be used as building blocks. This approach, commonly referred to as a junction-tree model, represents molecules as trees of chemically valid scaffolds, ensuring that intermediate and generated structures remain chemically consistent by construction \cite{jin2018junction}. Additionally, we envision that the proposed architecture can also be generalized beyond the current limitation of generating a single organic ligand for an otherwise fixed coordination geometry. The next natural step forward would involve encoding all ligands, coordination number and coordination geometry into the latent space and achieve a full, unconstrained generation of coordination compounds and their properties.

In conclusion, we have shown that semi-supervised learning and training-by-proxy make it possible to use variational autoencoders to explore the chemical space of complex molecules such as SMMs and generate novel coordination compounds with target magnetic anisotropy starting from datasets smaller than 1k CASSCF calculations. These results provide a proof-of-concept that generative models can be used to tackle complex chemical problems in an efficient way, thus extending the remit of such promising machine learning methods in chemistry.

\section{Methods}
\label{sec:methods}
\textbf{VAE data preparation and analysis.} 
In each training run, the SMILES strings in the training set were screened to determine the maximum sequence length and the full set of characters present. Using one-hot encoding, and padding shorter strings with a space character that was also included in the character dictionary, we generated a 2D tensor of dimensions (maximum sequence length $\times$ total character count) for each ligand. These tensors were then fed into the GRU layers of the encoder. 

After training, the generated strings are validated using RDKit and compared to the training set SMILES for identifying novel molecules. Since a single molecule can be represented by multiple different SMILES strings, we compare the canonical SMILES of each generated molecule, computed using RDKit, against those in the training set to determine novelty. Canonical SMILES are standardized string representations produced through a set of rules based on molecular composition and connectivity, ensuring a unique and reproducible representation for each molecule. We note that when generating the canonical SMILES, we first kekulized the strings, which removes aromaticity and stereochemistry for convenience, without affecting the model’s performance.

\textbf{VAE model structure.} 
The encoder consists of three layers of bidirectional GRUs with hidden dimensions of 128, 96 and 32, followed by two dense layers that learn the parameters of the latent probability distribution. The decoder, in turn, is composed of two bidirectional GRU layers (with hidden dimension of 128 and number of characters) and a final dense layer that predicts logits, which are then passed through a softmax activation function to enable robust, character-by-character reconstruction of data.
The DNN in the GAUSS model variant that performs a guided search of chemical space using structural information consists of three dense layers with dimensions 32, 8, and 7, respectively. The network uses Leaky ReLU and SiLU as activation functions and includes two dropout layers with a rate of 0.005.
In the second GAUSS variant, optimized on LoProps, the DNN comprises two dense layers with dimensions 32 and 4. This configuration employs a Leaky ReLU activation function and a dropout layer with a rate of 0.2.
Additionally, the model is best optimized when using a normalized KL loss with a weighting factor \(\alpha=0.01\) leading to the overall loss: \(\mathrm{Loss} = \mathrm{Loss}_{\rm recon} + \alpha \mathrm{Loss}_{\rm KL} + \mathrm{Loss}_{\rm DNN} \).

\textbf{VAE model training.} 
All machine learning models were developed within the PyTorch framework \cite{paszke2019pytorch}. The models were trained for as many epochs as required, until the improvement in loss fell below 0.001 over a window of 10 epochs. In most cases, convergence was reached within 70 training epochs. Data were passed to the model using a DataLoader, with the dataset organized into batches of 100 samples and randomly reshuffled at each epoch. 
We initiated the training by Adam optimizer and setting the learning rate to $10^{-3}$. We employed a scheduled training scheme using the ReduceLROnPlateau scheduler in PyTorch, in which the learning rate is halved if the loss does not improve by $10^{-4}$ over two consecutive epochs. This resulted in smooth and stable loss curves. The model parameters corresponding to the lowest observed loss during each training were saved and used for subsequent analysis.

\textbf{Electronics structure simulations.} 
We use the quantum chemistry software ORCA 5 \cite{neese2020orca} to perform both DFT and state-averaged CASSCF calculations. Scalar relativistic effects are treated using the Douglas–Kroll–Hess (DKH) method, with picture-change effects included up to second order to account for DKH corrections in the spin–orbit coupling operator. For DFT we use the BP86-D3(BJ) functional, and the DKH-def2-TZVPP basis set for all elements except Dy, which is treated using the SARC2-DKH-QZVP basis set.\\
Due to the difficulty of capturing the right wavefunction of Dysprosium using DFT, the optimization is performed replacing the Dy(III) core with Y(III), which behaves similarly for electronic bond configurations, yet lacks the f-like electrons responsible for its magnetic properties. The compounds are optimized first under constraints, namely the five planar oxygen atoms being locked in plane and with perfect 72° angles between them, and afterwards under complete relaxation, to an SCF convergence point of $10^{-8}$au.
After the geometry optimization, and initial guess for the CASSCF calculation is produced using a DFT calculation on the ligands without the metal core, and a mixed DFT and CASSCF calculation on the isolated Dy(III) core, and combining their wavefunctions. Finally a CASSCF calculation including the 7 f-like orbitals and corresponding 9 electrons is performed on the complete Dy(III) compound, and all Kramer doublet energies $\Delta E_{0j}$ extracted after diagonalization of the SOC matrix through Quasi-Degenerate Perturbation Theory.

\textbf{LoProp evaluation.} 
LoProp evaluation at the DFT level is performed with the OpenMolcas package (version 25.02)  \cite{li2023openmolcas,aquilante2020modern,fdez2019openmolcas}. Kohn-Sham DFT calculations are carried out using PBE exchange-correlation functional and by employing the ANO-R2 atomic natural orbital basis \cite{zobel2019ano} for all the atoms. Localized properties are obtained using the LoProp (Local Properties) scheme implemented in OpenMolcas \cite{gagliardi2004local}. In this approach, the molecular electron density obtained from the converged SCF calculation is projected onto a localized minimal basis, allowing for a partitioning of the density into atomic contributions. From such partitioning, atom-centered multipole moments of the connecting atom of each ligand are derived and used within the GAUSS framework. The retained LoProp properties are the charge, dipole moment magnitude, and isotropic polarizability.     

\noindent
\textcolor{black}{\textbf{Author Contributions}}\\
A.L. proposed the project and supervised its execution. Z.K. developed the software and designed the models' architecture together with A.L. L.F. contributed to ab initio simulations and the development of machine learning descriptors for GAUSS-I. L.A.M. performed the calculation of LoProps. Z.K. analyzed the results and wrote the manuscript together with A.L.

\noindent
\textbf{Acknowledgements and Funding}\\
This project has received funding from the European Research Council (ERC) under the European Union’s Horizon 2020 research and innovation programme (grant agreement No. [948493]). Computational resources were provided by the Trinity College Research IT and the Irish Centre for High-End Computing (ICHEC).

\bibliographystyle{achemso}
\bibliography{bibliography}

\end{document}